\documentclass{iopjournal}

\usepackage[numbers,sort&compress]{natbib}
\usepackage{amsmath}
\usepackage{siunitx}

\DeclareSIUnit{\erg}{erg}
\DeclareSIUnit{\Mass}{\mathit{M}}
\DeclareSIUnit{\c}{\mathit{c}}
\DeclareSIUnit{\dyn}{dyn}
\DeclareSIUnit{\Gauss}{G}
\DeclareSIQualifier{\Sun}{\ensuremath{\odot}}
\DeclareSIQualifier{\disc}{disc}
\DeclareSIQualifier{\BH}{BH}
\DeclareSIQualifier{\NS}{NS}
\DeclareSIUnit\clight{\text{\ensuremath{c}}}
\DeclareSIUnit{\kms}{\kilo\metre\per\second}

\DeclareUnicodeCharacter{2299}{\odot}

\definecolor{darkperiwinkle}{RGB}{64, 64, 128}

\begin{document}

\articletype{Paper} 

\title{A Reproducible Black Hole-Neutron Star Merger Gallery Example for the Einstein Toolkit}

\author{Rahime Matur$^{1,*}$\orcid{0000-0001-9488-3817}, Beyhan Karakaş$^2$\orcid{0000-0002-3566-7329}, Roland Haas$^{3,4,5}$\orcid{0000-0003-1424-6178}, Ian Hawke$^1$\orcid{0000-0003-4805-0309}, Nils Andersson$^1$\orcid{0000-0001-8550-3843} and Steven R. Brandt$^6$\orcid{0000-0002-7979-2906}}

\affil{$^1$Mathematical Sciences and STAG Research Centre,  University of Southampton, Southampton SO17 1BJ, UK}

\affil{$^2$beyhannkarakas@gmail.com}

\affil{$^3$Department of Physics and Astronomy, University of British Columbia, Vancouver, British Columbia, Canada}

\affil{$^4$National Center for Supercomputing applications, University of Illinois, Urbana, Illinois, USA}

\affil{$^5$Department of Physics, University of Illinois, Urbana, Illinois, USA}

\affil{$^6$Center for Computation \& Technology, Louisiana State University, Baton Rouge, LA, United States of America }

\affil{$^*$Author to whom any correspondence should be addressed.}

\email{r.matur@soton.ac.uk}


\begin{abstract}
Black hole-neutron star mergers are important sources of gravitational waves and potential electromagnetic counterparts, but publicly available numerical relativity configurations for these systems remain limited. In this work, we present a fully reproducible black hole-neutron star merger simulation performed exclusively with official Einstein Toolkit thorns and configured to target the detected event \texttt{GW230529}. We evolve the system at three resolutions, with finest grid spacings of $162$, $222$, and $310$~m, and assess the numerical robustness of the resulting dynamics and gravitational wave signal. The entire setup, from initial data to a parameter file with some of the analysis scripts, is publicly released as a new Einstein Toolkit gallery example and is distributed as part of the Hypatia release, establishing a reference black hole-neutron star merger configuration within the Einstein Toolkit.

\end{abstract}

\section{Introduction}

Black hole-neutron star (BHNS) mergers contribute to our understanding of gravity in the strong field regime, the characterization of gravitational waves (GWs) and electromagnetic (EM) counterparts, the behaviour of matter at and above nuclear densities and how these properties are affected by the intrinsic and binary parameters. Following the first gravitational wave detection from compact binary mergers, \texttt{GW150914}, identified as a binary black hole merger~\citep{GW150914::2016}, and \texttt{GW170817}~\citep{GW170817}, identified as a binary neutron star (BNS) merger, the first confirmed BHNS mergers, \texttt{GW200105} and \texttt{GW200115} completed the classification of astrophysical compact binary merger scenarios~\cite{BHNSobservtions:2021}. Despite these detections, the only multi-messenger event remains the BNS merger \texttt{GW170817} and its EM counterpart~\texttt{AT2017gfo}~\cite{GW170817,Kasliwal::2017,LIGO_em:2017}.

BHNS mergers can follow two distinct evolutionary channels: either the neutron star (NS) plunges directly into the black hole (BH), yielding no or negligible ejecta and no accretion disc, or it is tidally disrupted, producing a tidal tail that ejects cold neutron-rich material while the remnant matter forms a disc around the BH. Whether tidal disruption occurs, and thus whether a BHNS system can power an EM counterpart, depends primarily on three parameters \cite{reviewfoucart,reviewkyutoku}: the mass ratio ($q = M_{\mathrm{BH}}/M_{\mathrm{NS}}$ where $M_{\mathrm{BH}}$ and $M_{\mathrm{NS}}$ are the mass of BH and NS, respectively), the BH spin, $a$, and the NS compactness, $C$. Disruption becomes increasingly likely for lower mass ratios, higher BH spins, and less compact NSs, making these regimes the key targets for multi-messenger BHNS observations. 

Current GW detectors are most sensitive to the inspiral phase, where the chirp mass $\mathcal{M}$ is the best constrained parameter. In BHNS systems, matter effects, such as the tidal deformability, typically remain unconstrained due to the low signal-to-noise ratio and relatively large mass ratios that suppress tidal signatures. Assuming low spin priors for the NS, $\chi_{\mathrm{NS}} < 0.05$, the measured parameters were $\mathcal{M} = 3.41 ^{+0.08}_{-0.07}$, $\chi_{\mathrm{eff}}= -0.01 ^{+0.08}_{-0.12}$, $q=4.7$ for $\texttt{GW200105}$ and $\mathcal{M} = 2.42 ^{+0.05}_{-0.07}$, $\chi_{\mathrm{eff}}= -0.14 ^{+0.17}_{-0.34}$, $q=4.2$ for \texttt{GW200115}~\citep{BHNSobservtions:2021}.

More recently, the \texttt{GW230529} event~\cite{Gw230529:2024A} demonstrated that BHNS binaries can include low mass BH with $M = 3.6^{+0.8}_{-1.2} M_{\odot}$. Combined with the inferred chirp mass $\mathcal{M}=1.91 ^{+0.04}_{-0.04}$, the mass ratio -- which is a key parameter governing tidal disruption, the resulting mass ejection and thus the expected multi-messenger signatures -- was estimated to be $q = 2.6$. Fully general relativistic simulations of this event~\cite{foucart230529, RM:2024sign, RM:2025spin} found that the ejecta mass could reach $10^{-4}$ to $10^{-3}M_{\odot}$ even for zero effective spin, moving with a velocity of $0.2 - 0.3\,c$, indicating the possibility of multi-messenger signatures. However, the poor sky localisation of the event made the detection of an EM counterpart infeasible~\cite{Gw230529:2024A}. Nevertheless, this event is highly significant because it suggests that low mass BHs may be common in BHNS binaries, and that, provided the other tidal disruption conditions are met, such systems could lead to a multi-messenger detection.

Understanding the nature of BHNS merger events requires numerical relativity, the only tool that can self-consistently link the GW signal to potential EM counterparts. With this goal, many groups~\citep{foucart230529, RM:2024sign, RM:2025spin, 2024PhRvD.110h3017I, 2025PhRvD.112f3027I, 2025PhRvD.111f4023T, 2025arXiv250700113G, 2026PhRvD.113b4031M, 2026arXiv260119405M} have recently focused on general relativistic (magneto)hydrodynamics simulations of these systems. The first fully general relativistic simulations of BHNS mergers were performed using \texttt{SACRA}~\citep{sacra1, sacra2}, \texttt{SpEC}~\cite{SpEC:web}, and \texttt{IllinoisGRMHD}~\citep{DelZanna:2002rv, Noble:2005gf,Zach_igm12015, Zach_igm22020, Zach_igm32023}, employing $\Gamma$-law equations of state in different studies~\cite{shibatafirst, duezfirst, firstetienne}. Even so, the literature on BHNS mergers remains limited, and further work is needed to improve our understanding of these systems, especially before the third generation GW detectors.

To ensure reproducibility, our earlier work~\cite{RM:2024sign} produced the first fully reproducible publicly available BHNS merger dataset~\cite{RM:2024sign, RM2024zenodo}. In that study, the hydrodynamical and spacetime evolution were carried out using \texttt{WhiskyTHC}~\cite{Radice_thc::2012, Radice_thc2::2014, Radice_thc3::2014, Radice_thc4::2015} and \texttt{CTGamma}~\cite{Pollney_ctgamma::2011}, which are based on the~\href{https://einsteintoolkit.org/} {\texttt{Einstein Toolkit}}~\citep{loffler_einstein_2012, zilhao_introduction_2013, EinsteinToolkit:web, Et:Hypatia:26} but are not part of it officially. In this present work, we perform a single simulation with three different resolutions using only official \texttt{Einstein Toolkit} thorns, namely \texttt{IllinoisGRMHD} and \texttt{McLachlan}~\cite{Brown:2008sb, Kranc:web, McLachlan:web}, with minimal modifications from our original work that used \texttt{WhiskyTHC$+$CTGamma} simulation setup. This study serves as a basis for extending the science capabilities of the \texttt{Einstein Toolkit}, and can be straightforwardly updated to work with finite-temperature, composition-dependent EoSs, including magnetic fields and weak interactions, all of which are available within \texttt{IllinoisGRMHD}. In this paper, we do not focus on ejecta and disc diagnostics, and instead limit our analysis to GWs, constraint violations, and the final BH properties--mass, spin, and coordinate kick--and their dependence on resolution, using three different resolutions.

This study improves the capability of the \texttt{Einstein Toolkit}, which already supports BHNS merger simulations, by introducing a new gallery example that uses a simple $\Gamma-$law equation of state with $\Gamma = 2$ and $K = 100$. It establishes a solid foundation for future BHNS merger studies.

The paper is organised as follows. In Sec.~\ref{sec:numericalmethods}, we describe the initial configurations and the numerical setup and analysis methods used in our simulations. In Sec.~\ref{sec:results}, we present the main results. Finally, in Sec.~\ref{sec:conclusion}, we summarise our findings and conclude the paper.

\section{Numerical setup and analysis}\label{sec:numericalmethods}

The initial data is generated using \texttt{FUKA}~\cite{Grandclement:2009ju, fuka} with a resolution of $(r, \theta, \phi) = (13, 13, 12)$. We simulate a single model with chirp mass, $\mathcal{M} = 1.91$, and effective spin, $\chi = 0$, consistent with \texttt{GW230529}, where the individual Arnowitt-Deser-Misner (ADM) masses are chosen to be $M_{\mathrm{BH}} = 3.6 M_{\odot}$ and $M_{\mathrm{NS}} = 1.4M_{\odot}$ for the BH and NS, respectively. The stars are modeled as non-spinning with an initial separation of $45 \, \mathrm{km}$ corresponding to approximately $1.5$ orbits before merger. The total initial angular momentum is $J_0 = 17.04 M_{\odot}^2$.

The hydrodynamical evolution is performed using \texttt{IllinoisGRMHD}~\citep{Zach_igm12015, Zach_igm22020, Zach_igm32023}, which employs a three-velocity formulation that differs from the Valencia formulation of general relativistic magnetohydrodynamics~\citep{Valencia::1991, Valencia::1997, Valencia::1999}. Reconstruction is performed using Piecewise Parabolic Method (PPM)~\citep{PPM} and fluxes are computed with an approximate Harten–Lax–van Leer (HLL) Riemann solver~\citep{HLL}. The NS is modeled as a perfect fluid using a Gamma-law EoS with $\Gamma = 2$ and $K = 100$, where $\Gamma$ and $K$ are the adiabatic index and the polytropic constant.

The spacetime evolution is performed using \texttt{ML\_CCZ4} within \texttt{McLachlan}~\cite{Brown:2008sb, Kranc:web, kranc2, McLachlan:web}, which employs the Conformal and Covariant Z4 \texttt{(CCZ4)} formulation of the Einstein Field Equations~\citep{Alic_ccz4, Alic_2013} to control the constraint violation. All simulations using $\mathrm{ML\_CCZ4}$ evolve the shift vector with the driver variable $B^i$. We set the shift–Gamma coefficient to $0.75$ and the damping coefficient to $0.2$. The constraint damping parameters are set to $\kappa_1 = 0.02$ and $\kappa_2 = 0.0$, following the choices adopted in our previous work~\cite{RM:2024sign, RM:2025spin} to obtain stable BHNS merger evolutions. In both the \texttt{Z4c} and \texttt{CCZ4} formulations, these coefficients parametrize the same underlying Z4 constraint-damping prescription. We use the \texttt{1+log} and \texttt{Gamma Driver} gauge conditions, with shift advection enabled, commonly referred to as the moving puncture method. These gauge choices are the standard and well tested in numerical relativity. We use radiative boundary conditions in all simulations.

The computational domain extends to $\sim1985 \, \mathrm{km}$ in each direction centered on the center of mass of the binary at $(x,y,z) = (0, 0, 0)$, and no symmetries are imposed. Adaptive mesh refinement (AMR) is performed using \texttt{Carpet}~\citep{Schnetter:2003rb, CarpetCode:web}, the AMR driver of the \texttt{Cactus}~\citep{allen_cactus_1999, goodale_cactus_2003, Cactuscode:web, Cactusprize:web} framework. We use seven refinement levels, with the finest level individually covering each star with a radius of $\sim 15\,\mathrm{km}$. The corresponding grid spacings on the finest level are $162 \, \mathrm{m}$, $222\, \mathrm{m}$, and $310\, \mathrm{m}$ for the high-resolution (HR), medium-resolution (MR), and low-resolution (LR) simulations, respectively. When changing the resolution, we modify only the grid spacing rather than increasing the number of refinement levels. The public gallery example corresponds to the MR setup. On the reference machine (DiRAC/Cosma8), it requires approximately $30$ hours on $2$ nodes using $256$ cores, corresponding to $\sim 7680$ CPU-hours, with a memory usage of $\sim 140$ GB.

Spatial derivatives of the spacetime variables are computed using fourth-order finite-difference methods, while the derivatives of the hydrodynamical equations are handled with finite-volume schemes. We apply fifth-order Kreiss-Oliger dissipation to the metric variables with a strength of $\epsilon = 0.1$ to suppress high-frequency numerical noise following~\cite{2009PhRvD..79d4024E}. It is also used to ensure the positivity of the metric components, as originally suggested in \cite{2009PhRvD..79d4024E}. Time integration is performed using the Method of Lines \texttt{(MoL)} with a fourth-order Runge-Kutta method and a Courant-Friedrichs-Lewy factor of $0.4$. 

The BH is tracked using \texttt{PunctureTracker}, which integrates the shift vector in time, and the NS is tracked using \texttt{VolumeIntegralGRMHD}. Simulations are performed using the current development version of the \texttt{Einstein Toolkit}, corresponding to the \textit{Hypatia} release.

The merger time, $t_{\mathrm{merger}}$, is defined as the time at which the GW strain of the dominant $(l,m) = (2,2)$ mode reaches its maximum. All simulations are evolved for at least $\sim10 \, \mathrm{ms}$ after merger. GW quantities are calculated at a radius of $\sim738 \, \mathrm{km} $ without extrapolation to infinity. The strain is calculated by double time integration of the Newman-Penrose curvature scalar $\Psi_4$ using fixed-frequency integration~\citep{reisswig_ffi}, see~\cite{eff_spin} for details of the strain and spectra calculations. The GW spectra is defined as
\begin{equation}
    h^{22}_{\mathrm{eff}}(f)
    = f \sqrt{\left(
    \lvert \tilde{h}^{22}_{+}(f) \rvert^2
    + \lvert \tilde{h}^{22}_{\times}(f) \rvert^2
    \right)/2},
\end{equation}
where $\tilde{h}^{22}_{+}$ and $\tilde{h}^{22}_{\times}$ denote the Fourier transforms of the plus and cross polarizations of gravitational waves, respectively. The GW spectra are compared to the sensitivity curves of the Advanced LIGO~\citep{Aasi_2015} for the fourth (04)~\citep{aligo_O4high} and fifth (05) observing runs~\citep{aligo_O5}, as well as the Einstein Telescope (ET)~\citep{etd_sensitivity, maggiore_science_2020}, for a distance of $200\, \mathrm{Mpc}$.

Apparent horizons are located using \texttt{AHFinderDirect}~\citep{ahfinderdirect1, ahfinderdirect2, ahfinderdirect3}. The remnant BH properties are computed using \texttt{QuasilocalMeasures}~\citep{QLM} and \texttt{AHFinderDirect}. We use the \texttt{PostCactus}~\citep{kastaun_postcactus} library for the analysis of GW strain and spectra, radiated energy and angular momentum, the Hamiltonian constraint norm, the evolution of the mass and spin of the initial and remnant BHs, and the coordinate velocity of the remnant BH. However, the analysis tools used to create the gallery example are based on~\cite{VisIT}. The figures are generated using Matplotlib~\citep{Hunter:2007}.

\section{Results}\label{sec:results}

\begin{figure}
    
    \includegraphics[width=1.0\linewidth]{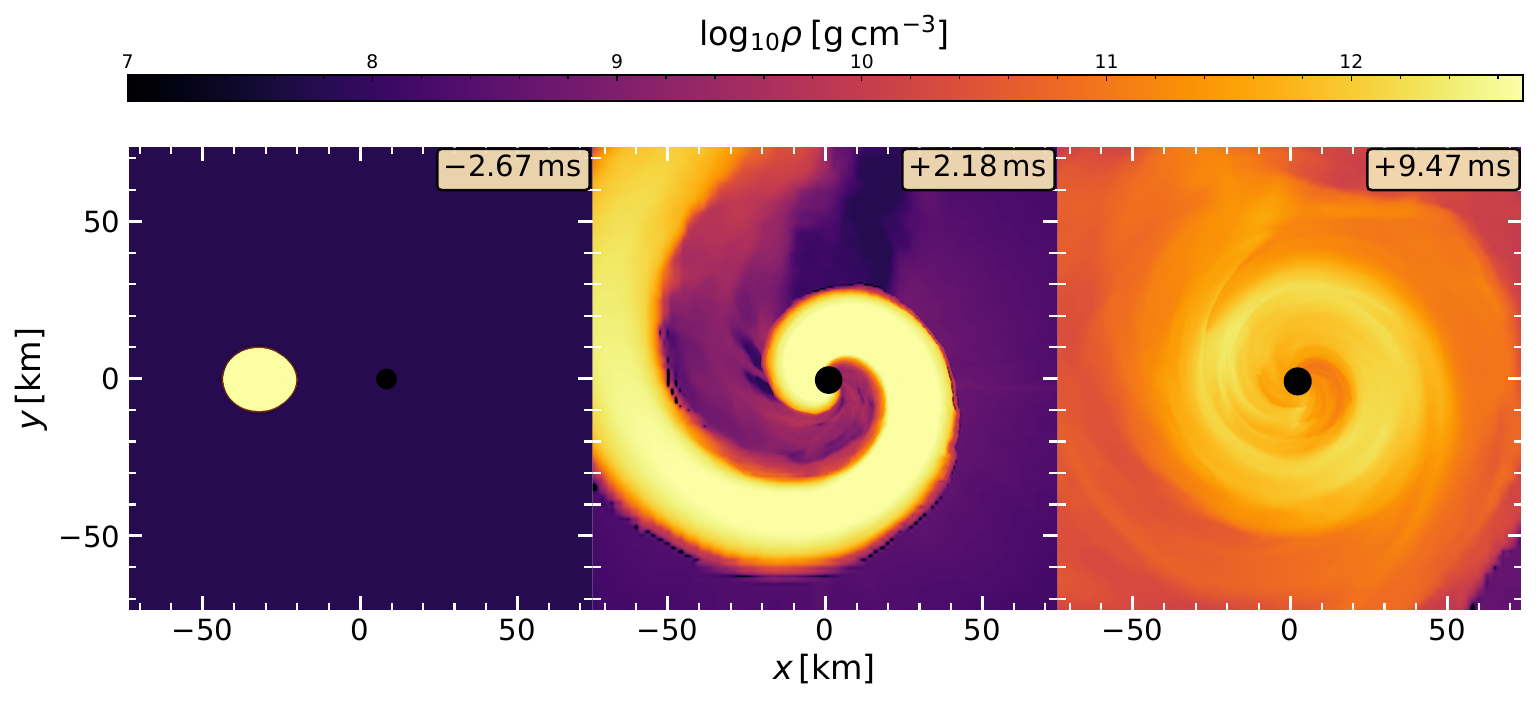}
    \caption{Snapshots of the rest-mass density on the equatorial plane from the medium resolution simulation before, during and after the merger. The panels show the system at the start of the simulation (left), during the merger when the NS is tidally disrupted (middle), and the remnant surrounded by a disc (right). Times shown are given relative to the merger time.}
    \label{fig:rho}
\end{figure}

\begin{figure}
    \centering
    \includegraphics[width=1.0\linewidth]{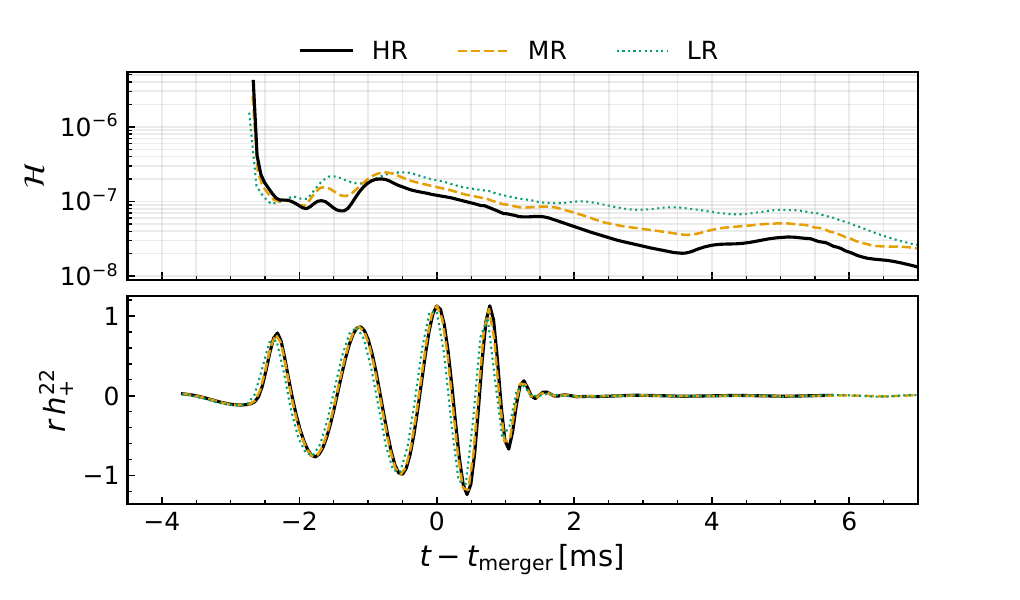}
    \caption{Volume-weighted root mean square norm of the Hamiltonian constraint (top panel), and GW strain extracted at $\sim738\, \mathrm{km}$ (bottom panel). The constraint violation decreases with increasing resolution, while the strain shows good agreement between HR and MR, with phase differences present at LR. The panels are aligned by shifting each simulation by its own merger time, and HR, MR, and LR represent high-, medium-, and low-resolution simulations, respectively.}
    \label{fig:ham_strain}
\end{figure}

\begin{figure}  
    \includegraphics[width=1.0\linewidth]{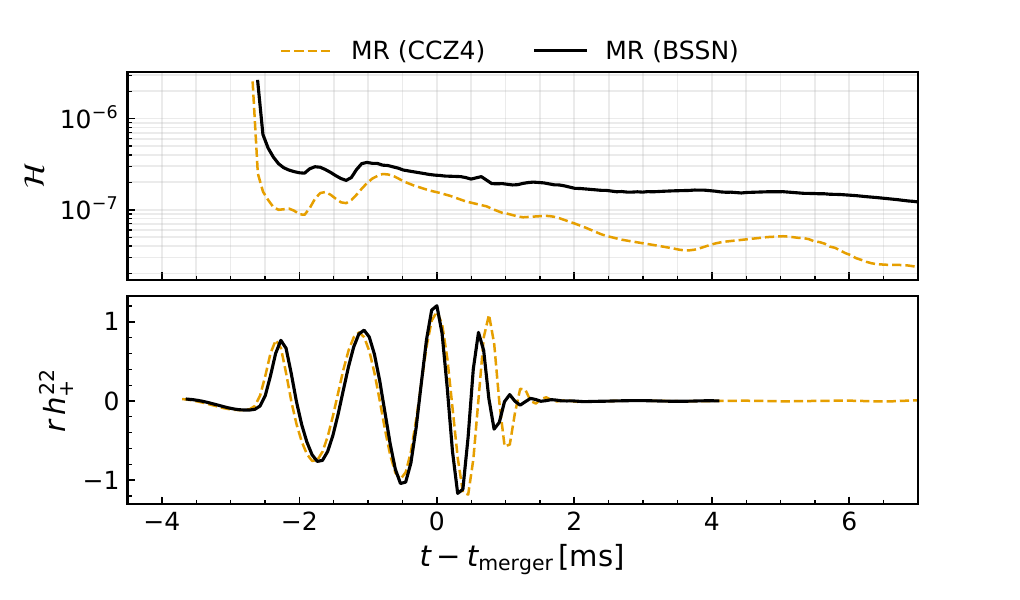}
    \caption{Same figure as Fig.~\ref{fig:ham_strain}, but now comparing BSSN and CCZ4 at medium resolution. CCZ4 exhibits lower constraint violations and only a very small phase deviation during the inspiral, whereas the phase difference increases in the post-merger phase.}
    \label{fig:formalisms}
\end{figure}

\begin{figure}  
    \includegraphics[width=1.0\linewidth]{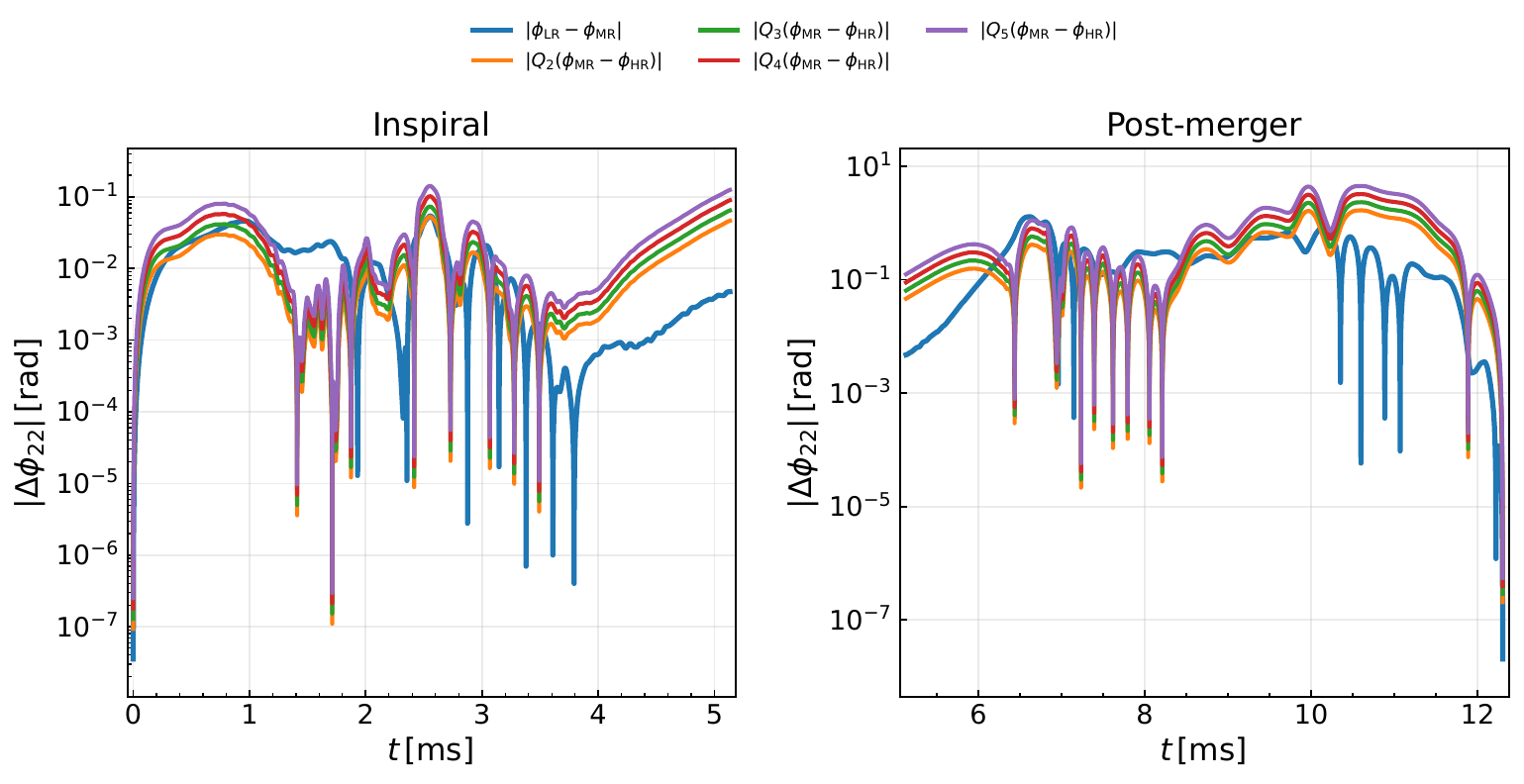}
    \caption{Convergence analysis of the GW phase using second-, third-, fourth-, and fifth-order convergence factors. Here, $\phi_{\rm LR}$, $\phi_{\rm MR}$, and $\phi_{\rm HR}$ denote the GW phase at low, medium, and high resolutions, respectively, while $Q_{2}$, $Q_{3}$, $Q_{4}$, and $Q_{5}$ represent the corresponding second-, third-, fourth-, and fifth-order convergence factors.}
    \label{fig:phasecon}
\end{figure}

\begin{figure}  
    \includegraphics[width=1.0\linewidth]{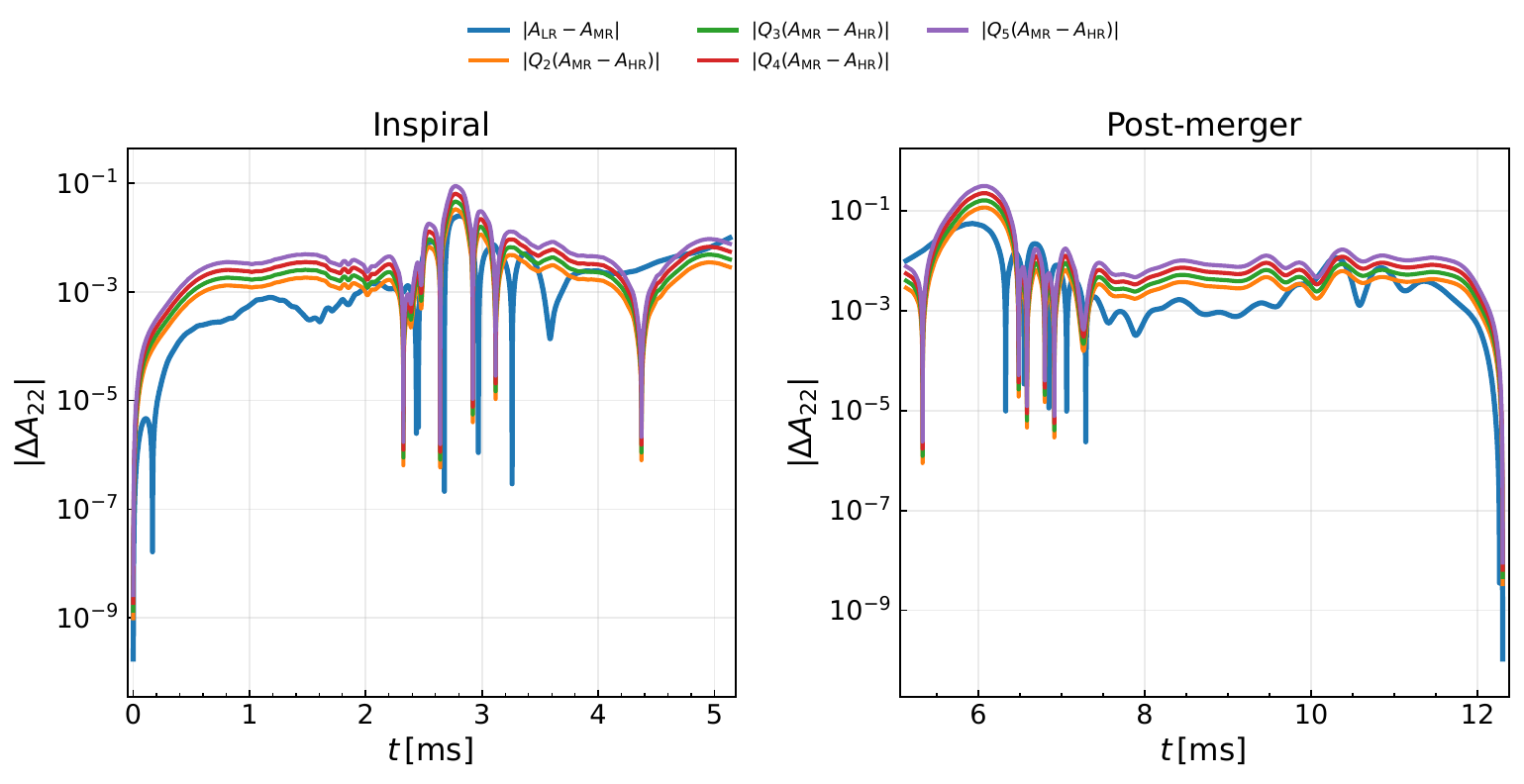}
    \caption{The same as Fig.\ref{fig:phasecon}, but showing the GW amplitude instead of the phase.}
    \label{fig:ampcon}
\end{figure}

\begin{figure}
    \centering
    \includegraphics[width=1.0\linewidth]{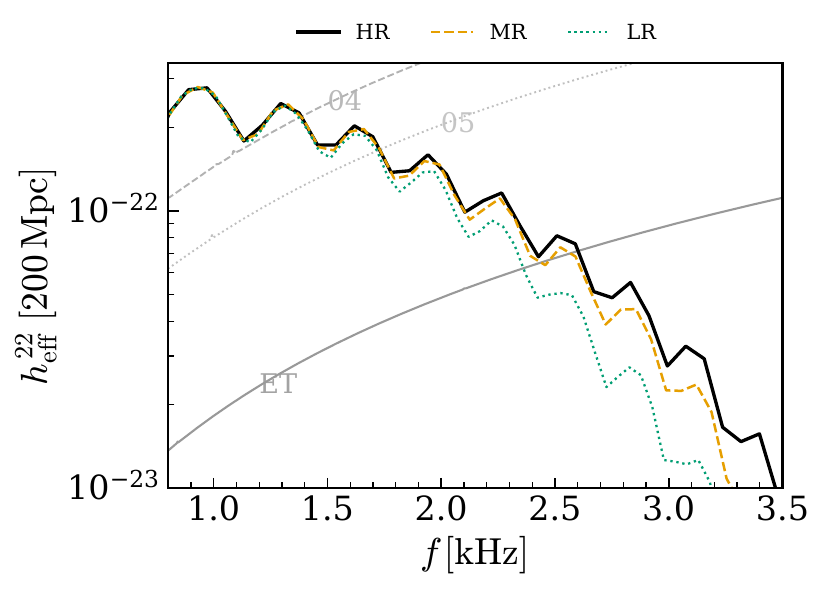}
    \caption{GW spectra for the $(l, m) = (2,2)$ mode at a distance of $200\, \mathrm{Mpc}$, compared to the sensitivity curves of the Advanced LIGO (04 and 05) and the Einstein Telescope (ET). HR, MR, and LR represent high-, medium-, and low-resolution simulations, respectively.} 
    \label{fig:spectra}
\end{figure}

\begin{figure}
    \centering
    \includegraphics[width=1.0\linewidth]{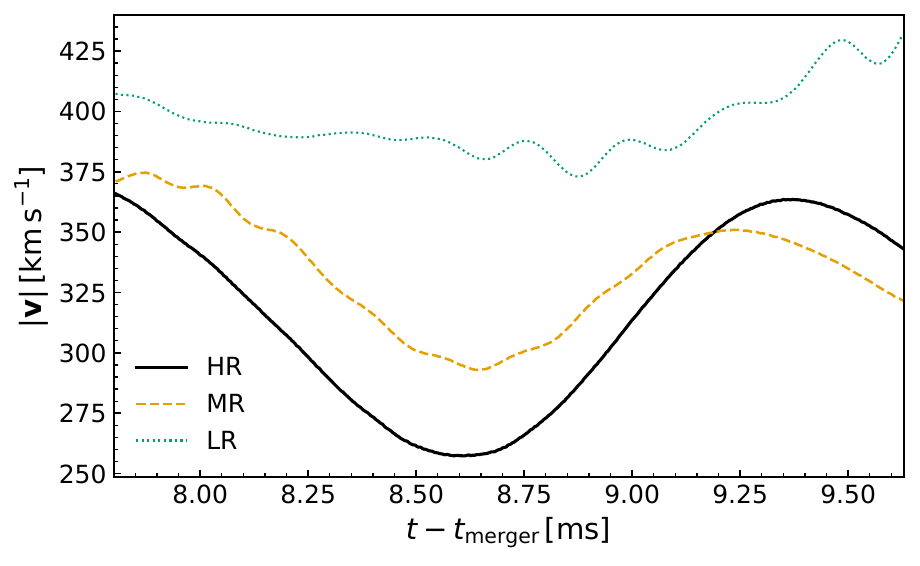}
    \caption{Evolution of the coordinate velocity of the remnant during the final $\sim2 \,\mathrm{ms}$. HR, MR, and LR represent high-, medium-, and low-resolution simulations, respectively. The oscillations are the result of residual gauge dynamics and should not be interpreted as physical.}
    \label{fig:vel}
\end{figure}

\begin{figure}
    \centering
    \includegraphics[width=1.0\linewidth]{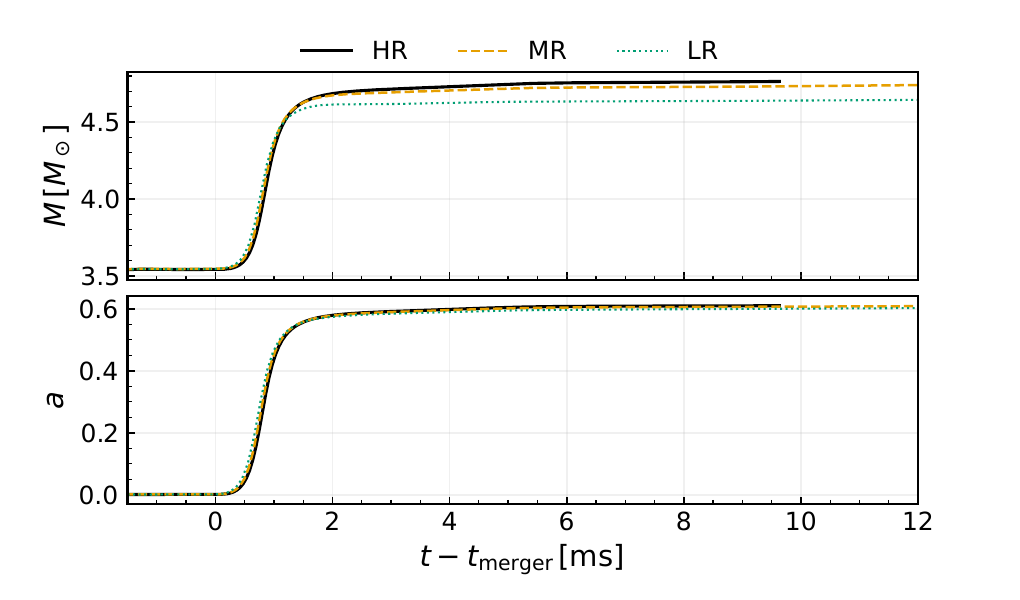}
    \caption{Quasi-local mass and spin evolution of the BH for high- (HR), medium- (MR), and low-resolution (LR) simulations. The remnant mass and spin are consistent between HR and MR, whereas the LR underestimates the remnant mass.} 
    \label{fig:qlm}
\end{figure}

In this section, we provide an overview of the GW signal, the rest-mass density, and the orbital dynamics. Figures~\ref{fig:rho}, \ref{fig:ham_strain}, \ref{fig:formalisms}, \ref{fig:phasecon}, \ref{fig:ampcon}, \ref{fig:spectra}, \ref{fig:vel}, and Fig.\ref{fig:qlm} show the rest-mass density on the equatorial plane, the evolution of the constraint violations and GW strain, the comparison between Baumgarte-Shapiro-Shibata-Nakamura (BSSN)~\citep{bssn1, bssn2, bssn3} and CCZ4 evolutions, convergence of the GW phase and amplitude, the GW spectra, the recoil velocity for all three resolutions, and the evolution of the remnant’s spin and mass. We report the total energy and angular momentum emitted in GWs, their fractions relative to the initial mass and angular momentum of the binary, and the remnant BH properties--including mass, dimensionless spin, and coordinate velocity--in Table~\ref{tab:gw_bh}.

We show three rest-mass density snapshots in Fig.~\ref{fig:rho}. As illustrated in the figure, the NS undergoes tidal disruption and leaves a non-negligible amount of matter surrounding the black hole, even though the BH is non-spinning. This behaviour is primarily driven by the relatively small mass ratio of the system, and the degree of tidal disruption is expected to increase further for higher BH spins~\citep{RM:2025spin}. As a consequence of the tidal disruption, the system is expected to produce electromagnetic counterparts in addition to the GW signal.

It is important to keep the constraint violations low to ensure a stable spacetime evolution. As shown in the top panel of Fig.~\ref{fig:ham_strain}, our simulations exhibit very small violations on the order of $10^{-6}$, which decrease to $10^{-7}$ within $2$ ms after merger. Although higher resolution further suppresses the violations, the LR run already performs well. This behaviour is consistent with the constraint control structure of the \texttt{CCZ4} evolution, which evolves the Z4 constraint variables and includes explicit constraint damping terms, neither of which is present in the standard \texttt{BSSN} evolution used for comparison (see Sec.~\ref{sec:numericalmethods}). In practice, we observe at least an order-of-magnitude reduction in violations compared to BSSN formulation (see Fig.~\ref{fig:formalisms}).

The bottom panel of the same figure shows that the system merges after completing only one and a half orbits. Following the merger, the BH rapidly settles into a stationary state, and the GW signal correspondingly flattens on a very short timescale of order $\sim 1$ ms. The effect of resolution appears as a small phase shift by eye, which we illustrate through the corresponding GW spectra. We present the GW spectra alongside the sensitivity curves of the current (O4) and future (O5) observing runs of Advanced LIGO and of the planned Einstein Telescope in Fig.~\ref{fig:spectra}. This comparison is qualitative and does not imply detector level distinguishability.

\begin{table}
\caption{Energy $E_{\mathrm{GW}}$ and angular momentum $J_{\mathrm{GW}}$ radiated in GWs, and their ratios to initial total mass $M_{\mathrm{tot}}$ and angular momentum $J_0$. Here $h$ denotes the resolution on the finest grid. $M$, $a$ and $|v|$ represent the mass, dimensionless spin, and coordinate velocity of the remnant at $\sim10\, \mathrm{ms}$ after the merger.}

\label{tab:gw_bh}
\centering
\begin{tabular}{c c c c c c c c}
\hline
$h [\mathrm{m}]$ &
$E_{\mathrm{GW}}\, [10^{52}\, \mathrm{erg}] $&
$E_{\mathrm{GW}}/M_{\mathrm{tot}}$ [\%] &
$J_{\mathrm{GW}}$ [$M_\odot^2$] &
$J_{\mathrm{GW}}/J_0$ [\%] &
$M$[$M_\odot$] &
$a$ &
$|v| [\mathrm{km/s}]$ \\
\hline
310 & 5.77 & 0.65 & 1.77 & 10.40 & 4.64 & 0.60 & 434 \\
222 & 6.21 & 0.70 & 1.86 & 10.91 & 4.73 & 0.61 & 321 \\
162 & 6.44 & 0.73 & 1.91 & 11.18 & 4.76 & 0.61 & 343 \\
\hline
\end{tabular}
\end{table}

The kick velocity reported in Table \ref{tab:gw_bh} and Fig.~\ref{fig:vel} shows that the system experiences a modest recoil of $\sim 300$–$400\ \mathrm{km\,s^{-1}}$, consistent with values reported in the literature, for example in~\cite{2023PhRvD.107l3001H}. The kick velocity of the remnant BH shows a non-monotonic resolution dependence. The recoil arises from a combination of linear momentum carried away by GWs and the contribution from dynamically ejected matter. Following the approach of~\cite{subsolar}, we estimate the GW contribution using \texttt{surfinBH}~\cite{surfinbh}. Although the model is formally calibrated for binary black hole systems, it provides a useful reference for this configuration and yields an expected GW-induced kick of $\sim 143\ \mathrm{km\,s^{-1}}$. The larger total recoil may also receive a contribution from dynamical ejecta.

Finally, we compute the quasi-local mass and dimensionless spin of the remnant BH as reported in Tab.~\ref{tab:gw_bh} and Fig.~\ref{fig:qlm}. For the HR simulation, the final values are $4.76\,M_{\odot}$ and $0.61$, respectively. While the resolution changes the mass very slightly, the spin parameter of the remnant is consistent across all resolutions. 

Since this study targets a configuration motivated by a real observed event \texttt{GW230529}, similar systems are likely to be detected in the future. In the era of third-generation ($3$G) GW detectors, such events are expected to become increasingly common, highlighting the need for systematic studies to fully characterise their merger dynamics and post-merger outcomes.

\section{Conclusion}\label{sec:conclusion}

In this paper, we present the results of a black hole-neutron star merger simulation using official \texttt{Einstein Toolkit} thorns. Our target system is motivated by the observed event \texttt{GW230529}. Based on this work, we also provide a new gallery example for the \texttt{Einstein Toolkit}, designed to support reproducibility. The parameter file, initial data, thornlist, supplementary data, and step-by-step instructions for compiling and running the simulation are publicly available through the \texttt{Einstein Toolkit}. The example also provides scripts for analyzing the gravitational waves and rest-mass density evolution.

The results include the gravitational wave signal and its detectability, the recoil velocity of the final black hole, final properties of the remnant, and snapshots of the rest-mass density. The measured recoil is in agreement with fitting formulas for binary black hole mergers, with an additional contribution likely arising from mass ejection. The Hamiltonian constraint violation remains under control throughout the evolution, reaching the order of $10^{-7}$ within $2$ ms after merger. 

In this study, we present the convergence analysis of the GW phase in Fig.~\ref{fig:phasecon}. During the inspiral, the phase differences show an overall convergence trend, although no single convergence order describes the entire evolution. Around and after merger, the differences no longer exhibit a clear convergence behaviour. In particular, during the first approximately $2$ms after merger, the differences grow and the curves progressively depart from one another.

The GW amplitude, shown in Fig.~\ref{fig:ampcon}, similarly exhibits a convergence trend during the inspiral. This behaviour persists approximately up to merger, after which the differences no longer scale consistently and no clear convergence order can be identified.
Although \texttt{IllinoisGRMHD} is formally second order in the hydrodynamics, the reported convergence order refers to GW phase and amplitude until the merger, not to the full GRHD evolution. These results demonstrate that reliable black hole--neutron star merger simulations can be performed using the presented gallery example.

\ack{This work used the DiRAC Memory Intensive service (Cosma8) at Durham University via a discretionary allocation (project ID dp414), managed by the Institute for Computational Cosmology on behalf of the STFC DiRAC HPC Facility (www.dirac.ac.uk). The DiRAC service at Durham was funded by BEIS, UKRI and STFC capital funding, Durham University and STFC operations grants. DiRAC is part of the UKRI Digital Research Infrastructure. RM also acknowledges the use of the IRIDIS High Performance Computing Facility, and associated support services at the University of Southampton, in the completion of this work. NA and IH also gratefully acknowledge support from the Science and Technology Facility Council (STFC) via grant numbers ST/R00045X/1 and ST/V000551/1.}

\data{The data, including the initial data and parameter file for the medium-resolution simulation, together with visualizations of $\Psi_4$ and the rest-mass density, are available as part of the new \href{https://einsteintoolkit.org/gallery/bhns/index.html}{black hole--neutron star merger gallery example} for the \texttt{Einstein Toolkit}.}

\suppdata{Supplementary data that consist of all raw data for the medium-resolution simulation is available on Zenodo~\cite{matur_2026_18940220}.}

\bibliographystyle{iopart-num}
\bibliography{references}

\end{document}